\documentclass[reprint,amsmath,amssymb,aps,twocolumn]{revtex4}
\usepackage{graphicx}% Include figure files
\usepackage{dcolumn}% Align table columns on decimal point
\usepackage{bm}% bold math
\usepackage[english]{babel}
\begin{document}

\title{Motion of a classical object with oscillating mass}
\author{Tomasz Lanczewski}
\email{tomasz.lanczewski@ifj.edu.pl}
\affiliation{H. Niewodnicza\'nski Institute of Nuclear Physics \\ Polish Academy of Sciences\\
    Radzikowskiego 152, 31-342 Krak\'ow, Poland}

\begin{abstract}
  We examine the problem of motion of an oscillating mass object provided no external force is applied to it. Calculations directly lead to the conclusion that the body accelerates in each system of reference except for the rest reference frame. Although its motion is periodic, it bears no resemblance to harmonic oscillations.
  \begin{description}
    \item[PACS numbers] 45.20.D-
  \end{description}
\end{abstract}

\maketitle
\section{Introduction}
  It is commonly known that, according to the First Law of Dynamics, in the absence of an external force, every object remains in rest or moves along a straight line with uniform velocity. The aforementioned law is valid under the assumption that the mass of the object in question is constant. In many textbooks (see e.g.~\cite{1}) and dedicated papers (see e.g.~\cite{2}) the problem of variable mass bodies or systems was thoroughly analyzed. However, in all the examples it is assumed that an influx or efflux of mass is a uniformly changing function of time. In order to fill the gap we focus here on the dynamics of an oscillating mass object. It turns out that the study of behavior of such a body brings forward interesting physical results and consequences.
\section{Equations of Motion of a Variable Mass Object}
  Let us consider an object with an initial mass of $m_{0}$ and its mass is some function of time $m=m(t)$. Additionally, we assume that there is no external force acting on the body ($\mathbf{F}=0$). The equation of motion in this case can be written down in the form
    \begin{equation}\label{p}
        \mathbf{F}=\frac{d\mathbf{p}(t)}{dt}=0,
    \end{equation}
  from which it clearly follows that $\mathbf{p}(t)=\mathbf{p}_{0}=const$.
  Now, since mass varies with time, we find that
    \begin{equation}\label{mfv}
        \mathbf{p}_{0}=m(t)\mathbf{v}(t).
    \end{equation}
   Substituting Eq.~(\ref{mfv}) into (\ref{p}) we get that
    \begin{equation}\label{Fgen0}
        \frac{dm(t)}{dt}\mathbf{v}(t)+m(t)\frac{d\mathbf{v}(t)}{dt}=0.
    \end{equation}
  The product in the second term of LHS of Eq.~(\ref{Fgen0}) may be equal to zero only if the first term vanishes. Under the initial assumption that the mass changes in time, it happens only if $\mathbf{v}(t)=0$.
  Rearranging the terms in Eq.~(\ref{Fgen0}) we immediately arrive at the equation
    \begin{equation}\label{a}
        \frac{d\mathbf{v}(t)}{dt}\equiv\mathbf{a}(t)=-\frac{dm(t)}{dt}\frac{\mathbf{v}(t)}{m(t)}=-\dot{m}(t)\frac{\mathbf{v}(t)}{m(t)}
    \end{equation}
  or if we utilize Eq.~(\ref{mfv}) we get
    \begin{equation}\label{af}
        \mathbf{a}(t)=-\frac{\dot{m}(t)}{(m(t))^{2}}\mathbf{p}_{0}.
    \end{equation}
  As we can see, in case $m=const$ we observe no acceleration and therefore we obtain the classical equation of motion of a body upon which no external force is acting ($\mathbf{a}=0$). However, in all other cases the acceleration turns up that is proportional to the mass change ratio. It also should be stressed here that, in general, the acceleration $\mathbf{a}(t)$ is not constant in time.

\section{Equations of Motion of an Oscillating Mass Object}
  In this section we deal with a massive object whose mass oscillates in time. The choice of such a mass function is due to a sound assumption that energy should be conserved in time, i.e. an influx and efflux of energy are then compensated. It also means that the mass of the body remains constant if averaged over an integer number of oscillations.

  Here we assume the mass function to be of the form
  \begin{equation}\label{mf}
    m(t)=m_{0}+\Delta m\sin\omega t,
  \end{equation}
  where $\Delta m$ is an amplitude of oscillations and $m_{0}\gg\Delta m$, and hence
  \begin{equation}\label{mfder}
    \dot{m}(t)=\omega\Delta m\cos\omega t.
  \end{equation}
  Substituting Eqs.~(\ref{mf}) and (\ref{mfder}) into (\ref{af}) we get that
  \begin{equation}\label{acc}
    \mathbf{a}(t)=-\frac{\omega\Delta m\cos\omega t}{(m_{0}+\Delta m\sin\omega t)^{2}}\mathbf{p}_{0}.
  \end{equation}\newline
   \begin{figure}[h]
   \centering
    %\begin{center}
    \includegraphics[width=0.5\textwidth]{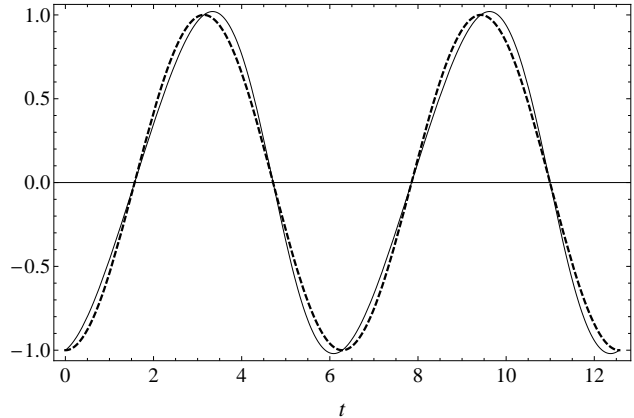}
    \caption{Acceleration 10$\bm{a}(t)$ given by Eq.~(\ref{acc}) (solid) vs.~$-\cos(\omega t)$ function (dashed). $\Delta m=0.1m_{0}$, $m_{0}=1$, $|\bm{p}_{0}|=1$ and $\omega=1$ were assumed here.}
    \label{fig1}
    %\end{center}
  \end{figure}
  As we can see in Fig.~\ref{fig1}, the acceleration function given by Eq.~(\ref{acc}) is significantly similar to a cosine function, and therefore resembles the oscillatory motion in which acceleration of a body is described by such a function.\newline
  By integrating Eq.~(\ref{mfv}) in which the mass function $m(t)$ is given by Eq.~(\ref{mf}) we get
  the solution of the equation of motion in the form \cite{2}
  \begin{eqnarray}\label{xft}
    \mathbf{x}(t) & = & \mathbf{p}_{0}\int\frac{dt}{m_{0}+\Delta m\sin\omega t} \nonumber \\
                  & = & \mathbf{p}_{0}\frac{2}{\omega\sqrt{m_{0}^{2}-(\Delta m)^{2}}}\arctan\frac{m_{0}\tan\frac{1}{2}\omega t+\Delta m}{\sqrt{m_{0}^{2}-(\Delta m)^{2}}}.\nonumber \\
  \end{eqnarray}
  The function described by Eq.~(\ref{xft}) is plotted in Fig.~\ref{fig2} for the same initial conditions and values as in Fig.~\ref{fig1}.
  It is clearly noticeable that this type of motion is periodic, however it bears no resemblance to the regular oscillatory motion. It can also be easily checked that for $\Delta m=0$ Eq.~(\ref{xft}) turns into well-known formula $\mathbf{x}(t)=\mathbf{v}t$.

\section{Conclusions}
  As we have shown, in the absence of an external force applied to a body it is possible that the body will accelerate provided its mass changes with time. Except for the rest reference frame, an observer notices non-zero acceleration of the body. The example directly indicates that the acceleration of an oscillating mass body is similar to the case of a classical oscillator, however the solution of the equation of motion reveals a significant difference from the oscillatory motion. These results may be important for many-body systems in which mass is exchanged between particles or bodies. In view of the aforementioned facts it may also be a contribution to a comprehension of quantum null oscillations on a classical ground by applying the Heisenberg uncertainty principle $\Delta E\Delta t\geqslant\frac{\hbar}{2}$ which states that the energy (and mass) of the system is not \emph{strictly} conserved and therefore serve as an attempt to establish a common ground for classical and quantum mechanics.
  \begin{figure}[ht]
    \centering
    %\begin{center}
    \includegraphics[width=0.5\textwidth]{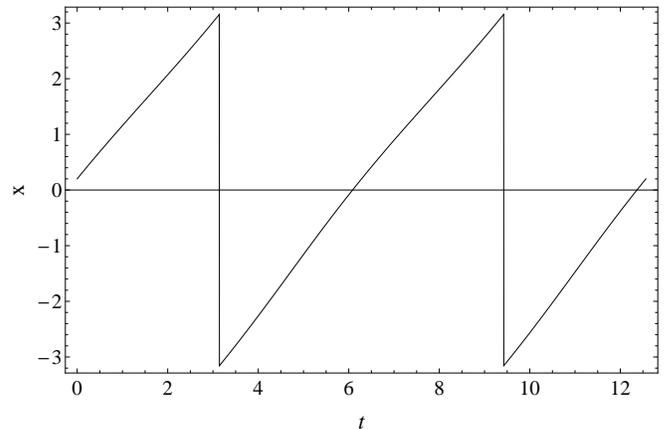}
    \caption{Solution of the equation of motion given by Eq.~(\ref{xft}). $\Delta m=0.1m_{0}$, $m_{0}=1$, $|\bm{p}_{0}|=1$ and $\omega=1$ were assumed here.}
    \label{fig2}
    %\end{center}
  \end{figure}

\end{document}